# A New Framework to Adopt Multidimensional Databases for Organizational Information System Strategies

Bryar A. Hassan*, Shko M. Qader**

* MSc Lecturer and Researcher, Department of Information Technology, Kurdistan Institution for Strategic Studies and Scientific Research, Sulaimani, Iraqi Kurdistan

bryar.hassan@kissr.edu.krd

* Consultant and IT Manager, Sulaimani Governorate, Sulaimani, Iraqi Kurdistan

shko.qader@suligov.com

## Abstract

As information becomes increasingly sizable for organizations to maintain the challenge of organizing data still remains. More importantly, the on-going process of analysing incoming data occurs on a continual basis and organizations should employ existing procedures that may not be adequate or efficient when attempting to access specific information to analyse.

In these latter days of technological advancement, organizations can offer their customers extensive data resources to utilize and thus accomplish individual objectives and maintain competitiveness; however, it remains a challenge in providing data in a format that serves each client's suited needs. For some, the complexity of a data model can be overwhelming to utilize. Furthermore, companies should secure an understanding of the purchasing power used by specific consumer groups to remain competitive and ease the operation of data analysis. This research paper is to examine the use of multi-dimensional models within a business environment and how it may provide customers and managers with generating queries that will provide accurate and relevant data for effective analysis. It also provides a new framework that can aid various types of organisations using sizable database systems to create their own multidimensional model from relational databases and present the data in multidimensional views. It also defines the requirements. Despite the availability of set tools, the complexity of utilizing the conceptions discourages customers as they may become apprehensive about exploring these options for analytical purposes. This could be done by conducting a query, syntactically accessed returns may produce incorrect information. A key suggestion to the issue may be found by encapsulating a relational schema among defined business terms. In addition, the encapsulation of the query terminology and syntax are related to business articles. Moreover, the outcome is made possible by the business oriented procedure referred to as the Online Analytical Process (OLAP).

# 1. THEROY AND BACKGROUND

Over the last decades, significant analysis has been made in the area of database technologies. Multidimensional database models are designed specifically to support the production of accurate and relevant data analysis in present time. According to (Pedersen and Jensen, 2005), multidimensional database technology includes three important aspects within data analysis. Firstly, it is used in data warehouse environments (Jensen et al., 2010) where data warehouses' consist of integrated data sources for specific aims of data analysis. Secondly, it lies at the core of the (OLAP) application. The OLAP application provides quick answers in queries of large quantities of data in order to determine the overall trends in an organisation. Thirdly, it is progressively becoming the basis for data mining. Moreover, data mining is the process of exploration and analysis by semiautomatic means to discover any unidentified information within large quantities of data (Kantardzic, 2011).

## 1.1 Multidimensional Databases Modelling

A multidimensional database (MDDB) is a specific kind of database that has been optimised for online analytical processing (OLAP) and data warehousing. Furthermore, it is designed specifically to support data analysis (Jensen et al., 2010). It is a technological solution that enables interactive analysis of huge amounts of data to be produced; also commonly used for analysis and data mining applications for decision making purposes (Pedersen and Jensen, 2005).Moreover, it is a type of databases for which data in multidimensional databases are saved in cells and the position of each cell is characterised by a number of hierarchical elements called dimensions. For example, a cell can represent a business event thus; the value of dimensions will indicate where and when the business event will take part. In addition to what has been said, MDDB is a relatively recent and popular phenomenon. In describing the MDDB, it is often been referred to as a hierarchy of aggregated values. When comparing hierarchical values, those that are higher contain additional aggregations in comparison to those that are lower in the hierarchy. The benefit of the hierarchical organisation provides users the option of effortlessly navigating among high and low accuracy visions of the same aggregated values by using drill-down to discover more detailed information and roll-up to perceive an overview. While drill-down enhances the precision of aggregate data, the roll-up by contrast is viewed as reducing the precisions of viewed data. In MDDB, users are able to pivot the desired data to view information from different directions.

## 1.2 Data Warehouse

Data warehouse is designed to assist in storing extensive amounts of data and support managers effectively in decision making. According to (Kimball, 1998) multidimensional database is a relational data warehouse, where the data is stored and organised in the Star Schema. For this reason, most of the data warehouses are considered to be decision support systems (DSS). A benefit to using the application of multidimensional models is that it can be utilised in data warehousing (Jensen et al., 2010). According to (El-Sappagh et al., 2011), building data warehouse may depend on three essential factors. These include data sourcing, destination areas and mapping area when mapping area forms part of extract, transform and load (ETL). For effective data analysis, information in the data warehouse may come from operational distributed online transaction processing (OLTP) sources into a single data warehouse for online analytical processing (OLAP). Interestingly, the term data warehouse was first introduced by W.H. Imnon (Inmon, 2000). Data warehouse can be defined by subject oriented. Data is then organised for themes (integrated) which implies that data is consistent in the database in accordance with an organisation's operational applications. Moreover, time variant is a time frequency component in the data warehouse where presenting data on the report can be modified over time. Furthermore, data referred to as non-volatile in the data warehouse is never over written or deleted and may be stored for further reporting. Arguably, the historical data in the data warehouse is organised, integrated and can be applied to support other relevant systems. The extract, transform and load (ETL) process can be used in data warehousing. Individually, extract is obtains data from external data source when most data warehousing projects usually combine data from variety source system. Moreover, transform is the process of conversion from the extracted data into another required database form. Finally, data is loaded into the end target database (data warehouse). On the other hand, it can be said that those process are able to convert databases from one type to other types(Gupta, 2013).

### 1.3 Online Analytical Processing (OLAP)

Online analytical processing (OLAP) utilises a multidimensional approach in order to manage and analyse historical data for operations and control purposes. Effectively, the data is collected and stored in highly optimised structures that will assist a business to explore the multidimensional data interactively from different perspectives (Nelson, 2008). In addition, it is characterised by queries which involve aggregations and could be very complex. There are three basic analytical operations are included in OLAP. First, roll up which means that the aggregation of data can be accumulated. Second, drill down is a procedure that allows consumer to navigate data through details. Third, slicing and dicing which allow users select a specific set of data from the OLAP cube and then view the slices from variety viewpoints (Gao et al., 2012) .

## 2. METHODOLOGY AND DESIGN

### 2.1 Overall Process of the Conceptual Framework

By incorporating elements of the relational database into the multidimensional model, a user can effectively access the desired data being requested to analytical purposes.
A multidimensional model consists of a fact table and several dimension tables. In the process of building a multidimensional model, the development of selected measures should occur. Initially, the user will need to select the desired measure components, either numerical property or formula that represent: product, quantity and/or the amount from the relational model. Further, each measure component is located within a column on the fact table; also regarded as a cube measure.

Moreover, measure attributes from the relational model will be saved into the fact table of the multidimensional model and will be regarded as the dependent attribute. After the measure attribute has been selected, requested criteria must be inputted as it relates to the desired component. Moreover, the purpose for the requested criteria allows the database to define the dimensions and provide the most accurate material successfully. In addition to what has been said, the grain of the model should be stated to identify the necessary information in the dimensional model. Furthermore, the information is selected from the relational model. Once the grain attribute is selected, the desired component must find a path between the measure attribute and the grain attribute. Moreover, the path chosen for the multidimensional model is important as whatever path is accepted will determine the theory of information to be used. It is important to note that both the grain attribute and path represent extensive pieces of data that are stored in the multidimensional model being adopted. Based on what grain and path are utilized, a new dimension will be created. Notably, a fact table will be defined by their grain (Dash and Agarwal, 2001). In a sales scenario, the (sales) fact table may express several items such as sales value by day by product name by store location (sales value = day-> product ->store location). The reason for depending fact table for both grain attributes and path is because of composed for those components. Each component in relation to the fact table, (grain attributes and the path) depends on one another to because they are composed of specific data. The process of the proposed framework consists of several steps as Illustrated in figure 2-1.

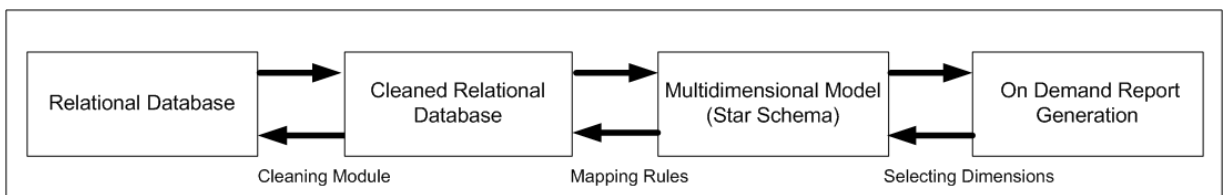

Figure 2-1: The Proposed Framework

## 2.2 Operational Data Source

Initially, the user of the framework will need to know which operational data sources are required. The operational source contains the operational data (Zaman et al., 2010, Agrusa et al., 2012). This information is important to determine the construction of the data model in the proposed multidimensional database. To further elaborate, the relational database must be established at the start of the operational source because the multidimensional model is organised for a relation in the database. Secondly, the operational source has to be in Third Normal Form (3NF) according to (Lechtenbörger and Vossen, 2003) for the reason that the model selects information in the normalised relational database. In addition to what has been said, normalisation is the process of organising attributes and tables efficiently in the relational database in order to minimise redundancy and dependency. Those conditions of the database are aided for the correct design of the multidimensional model. In this report, the relational source assumes that it is in the (3NF). As proof of this concept the Oracle database management system (DBMS in short) has been used as a data source and the data is physically stored in the Oracle server. DBMS is sometimes called a database manager because it allows the user to create and access data in a specified database. Moreover, in order to access the database server, essential information is requested. The information may consist of the following:

- The name of database schema.
- The Internet Protocol (IP) address of the database server.
- Server port number.
- User name and password of the user database manager.

Once the user is connected to the database server, the relational schema of the database should be extracted from the database manager and used for the purpose of designing the multidimensional database. Essentially, the purpose of the database schema is to characterize how the data is organised; utilization of the meta-data.

The meta-data refers to the collection of tables, attributions and relations between tables. The meta-data containing information about each table is drawn out. Thereafter, the information about the relational database is only necessary. However, the information is not ample thus far. It needs additional information about the attributes that are found within individual tables of the collection. The attribution information consists of the name and data type as well as the limitation of the attribute.

Interestingly, the name of the attribute is also the name of the column. Similarly, the attribute type is one of the standard SQL data types. Moreover, attribute limitations can include the unique, primary, foreign key, or none.

The meta-data of relations occurs between two different or duplicate tables and is sourced via foreign keys and primary keys in the attributes. In addition, each primary key attribute will be referenced in the foreign key or different tables. Moreover, the attributes maintain the same attribute data types.

## 2.3 Data Selection from the Operational Data Source

Data selection is one of the essential steps for creating a multidimensional model. It has been stated that initially, a user will chose all the desired attributes of the data source process without providing the information about what is being sought. This strategy may reflect the complexity of utilizing the existing operational data source. Thus, resulting in ramifications or discouragement to employ this method altogether. However, using measure selection mitigates the data selection process. The multidimensional database will incorporate the measure selection process to carry out the first step of choosing the attributes from the relational model. After selecting the measure attributes, one or more of the aggregation functions has to be applied on the measure attributes.

## 2.4 Building Multidimensional Model

It can be argued that the building of a dimensional data model is the most vital step for creating a dimensional database. The structure of the basic multidimensional model is:

$S = \{ F, D \}$
S represents an n-Dimensional fact schema.
F = represents Fact Table
D = represents $\{D_1, D_2, D_3, ....., D_n\}$ set of dimension tables.

According to (Pujolle et al., 2011), the top-down and bottom-up approaches are the two methodologies which can be used to build a multidimensional schemas in order to complete the dimensional database design. In building the dimensions, the methodology uses a top-down approach. Because the process in top-down approach starts the existing data source from an organisation's database. After the requirements are identified, a fact table can be built based on dimension tables and measure attributes. An outline of the methodology is listed:

- Identify the business process to analyse the requirements to be modelled.

- Determine the attributes of fact tables.
- Determine dimensions and hierarchies for each fact tables.
- Identify measure attributes for the fact table.
- Identify the attributes for each dimension tables.

Notably, a dimensional database model can have more than one fact tables. However, for the purpose of the proof of concept in this thesis one fact table will be utilized to simplify the technique.

## 2.4 Building a Hierarchy and Star Schema

The hierarchy represents different levels of grouped information that are organized by dimension tables. Notably, a hierarchy consists of an organized set of grain attributions and a set of attributes. Though attributes can be characterized through the levels of a hierarchy, a descriptive attribute is described by its entities. According to (Fernandez Ortega, 2013), the set of attributes within the hierarchy are interrelated through a series of levels that begin with the grain attribute. Moreover, hierarchies are identified by the drill up or drill down operations and will provide a level of detail relating to the dimension table. For example, the drill up operation provides an increased specified level of detail. Figure 2-2, presents both drill up and drill down operations.

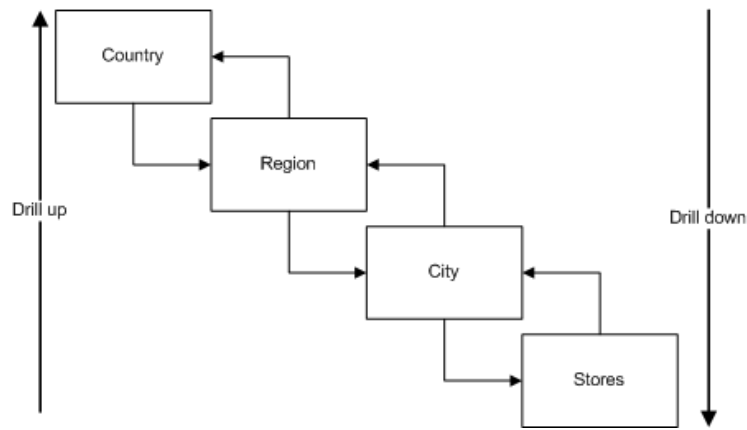

Figure 2-2: Example of drill up and drill down operation hierarchy

The paths of the hierarchy are determined by two factors. It begins in the grain attributes and it ends with the final table in the hierarchy by means of its primary key.
Furthermore, a star schema or dimensional model is composed of several elements. These elements include one or more fact tables, a single or set of dimension tables, and one-to-many join constraint relationships between the demoralised dimension tables and the fact tables. It is important to note that each row in a dimension table can describe many rows in the fact table. Furthermore, fact and dimension tables are at the core of a multidimensional database (Christian and Torben Bach, 2007).

## 3. IMPLEMENTATION AND CASE STUDIES

The implementation of the conceptual framework for this piece of work involves different steps in order to convert the relational database, a data source into one of the relational database presentations using either a star schema or snowflake schema as illustrated on figure 3-1. After some consideration, it is decidedly preferred to utilize a simple relational presentation which is a star schema. In addition, it was essential to use existing Oracle processes to complete the architecture. The relational database by Oracle database and a proposed tool to design the start schema by Oracle SQL developer have been chosen because it may allow connectivity to the database server and access to retrieve all information from a single source. In addition, the process for creating a new schema will be possible. As a result, the new schema provides information by the servers and database schema on the different database server.

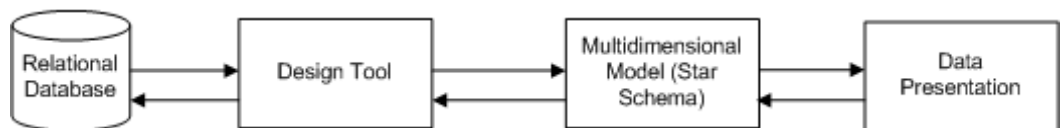

Figure 3-1: Conceptual Framework Implementation

Additionally, the proposed framework attempts to illustrate the process of designing star schema from existing relational databases. Figure 3-2, shows the client database tools have the ability to connect the database server while simultaneously creating another schema on the same server or another server. In realistic terms, a different database server has was preferential however, for the purpose of the proposed framework, the same server was been used but in the different schema.

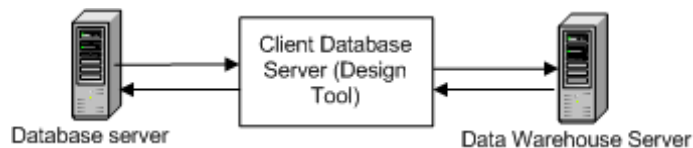

Figure 3-2: Client database server abilities

### 3.1 Implementation

Star Schema is arguably the most simplistic method in the data warehouse schema. According to (Karayannidis et al., 2002), the star schema of the relational database has numerous advantages such as: the star schema presentation is simple and easy to comprehend for query formulation, it can be easily accessed, and few number of joins are required to express queries because of the high level of demoralisation. As discussed before, the star schema consists of a fact table and a number of dimension tables. It is crucial to note the relationship between them is based on the one-to-many relationship. In other words, building the schema is based on the requirements of a business' needs. For instance, the manager of the given case study scenario may inquire about the following:

- What types of products are sold?
- Where the products are sold?
- Who is purchasing the products?
- When are products sold?

### 3.2 Case Studies

**4.2.1 Retail Case Study**

The retail case study scenario contains all the operational information that might support to illustrate the steps in the process of developing a proposed framework. In other words, this case study illustrates the hypothetical demands of retail stores.

Certainly, each organisation has a special demand on requesting information about their business. For that reasons, multidimensional database could be supported to view multiple data sets for clients who need to analyse the relationships between data categories. The clients could be for example financial analysts or marketing analysts.

More specifically, a marketing analyst might ask the following questions:

- What were the profits for Twinning's Green Tea for the (given) week? How does this product compare to sales in the same week over the last twelve months in the year? How did the product perform with stores in the area?
- Which store gets more customer traffic?
- How returns occurred for the tea product last week?

According to the relational database in organisations, the data views are restricted by both the database outline, and the structure that defines all elements of the database. By incorporating OLAP and using multidimensional databases, a marketing analyst is be able to pin the data to see information from various viewpoints, drill down and discover more detailed information, or dill up to see an overview of data.

*Requirements of Dimension Tables*

In order to process the request by the marketing analyst example above, dimensions must be built. Such dimensions are composed of one or more hierarchies. Furthermore, dimensional attributes assist to describe the dimensional vales. The attributes of the dimensions are descriptive or textual values that enable the system to answer marketing query.

*Requirement of fact Tables*

In addition to combining all foreign keys from dimension tables, measure attributes have been selected as key business vales. In this case, items sold on a daily basis will analysed both measurer (fact) attributes quantities and the total price. The measure is produced by Group by SQL aggregated functions. As a result, it can be seen that the fact table contains facts with the same level of aggregation.

As a result, figure 3.3 details the case study scenario multidimensional model from relational model, and figure 3-4 shows a data cube and multidimensional model logical view in retail case study scenario.

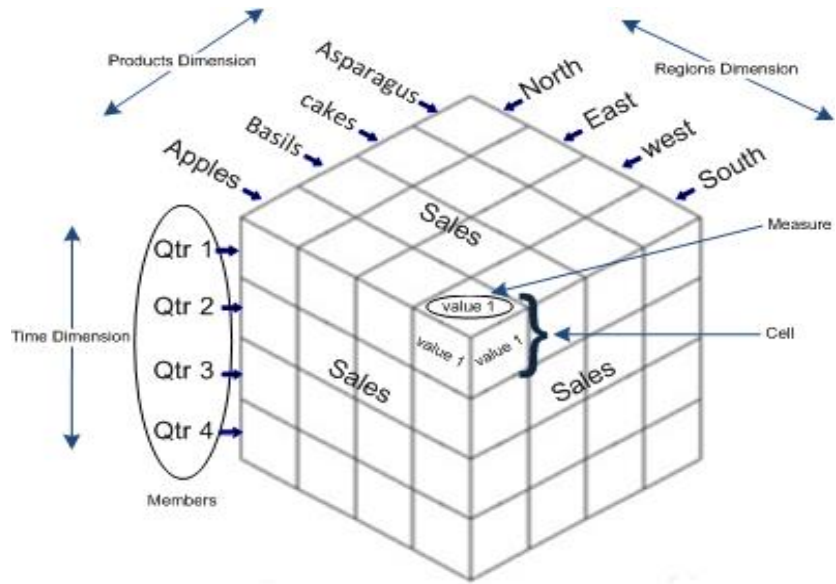

Figure 3-3: Sales Data Cube, multidimensional model logical view in retail case study scenario

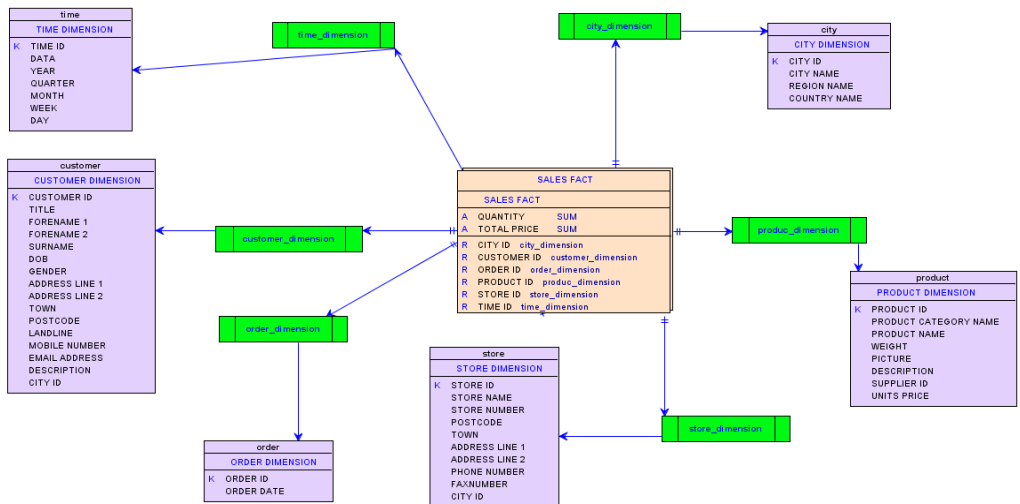

Figure 3.4: Retail case study scenario multidimensional model from relational model

### 3.2.2 Gorannet Company

The class diagram that created for the Gorannet Company conveys how subscribers are considered the most important entity type. Each subscriber is identified through their individual Social Security Number (SSN) and includes additional data such as Name, Age, and landline Number, with other attributes.

*Technology used*

The current version of the database application being used at Gorannet Company is Oracle version 11g. All departments are accessing the central database to manage their works. PHP script has been used to connect the database system in order to access the database.

*The problem*

The database transactions (OLTP) results in an extensive database because it saves all transaction data day-to-day information. Moreover, OTLP retrieves the data for generating daily, monthly and yearly reports. In some cases, SQL commands have many table joins in order to obtain the results for analytical purposes. Additionally, the data visualisation is problematic because viewing data is limited to only two dimensions.

### *Recommendation*

Multidimensional database provides the retrieval of data in an expeditious manner. By using a data warehouse application the database is optimised for retrieval and analysis. In this case, a new data warehouse server should be prepared for effective performance. There are four steps to creating a data warehouse. Figure 3-5, illustrated the process of design.

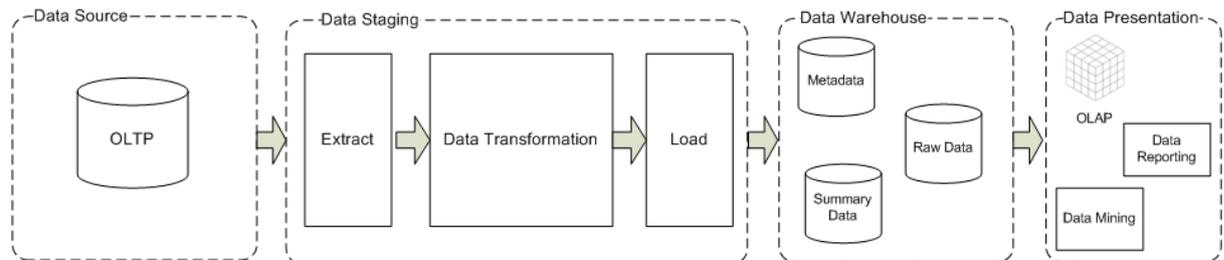

Figure 3-5: Data warehouse design

### *Identifying the fact and dimensions tables*

A data warehouse database is prepared once all the steps have been completed. Thereafter, a sample schema is created based on user requirements and preference of selected factors. Moreover, the data warehouse is influenced by the availability and structure of data in the operational systems. Figure 3-6 shows the overall steps of creating the physical schema in this situation.

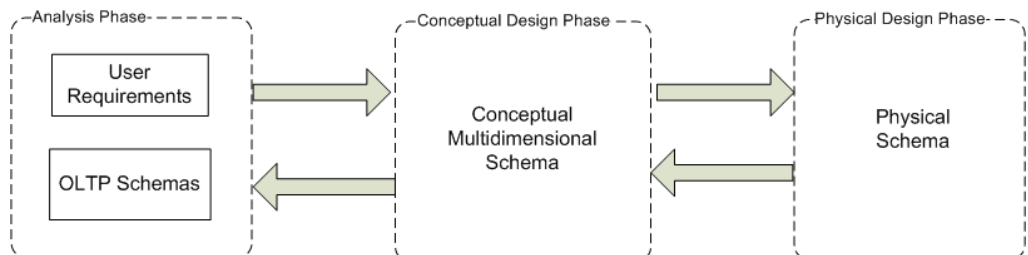

Figure 3-6: Schema Design Process

In other words, when creating the sample schema the following are required:
1- The number of subscribers.
2- Subscription time.
3- Location of using the Internet services.
4- Types of internet services.

Resultantly, figure 3-7 demonstrates the simple star diagram for the Gorannet Company, and figure 3-8 shows Gorannet Data Cube and multidimensional model logical view

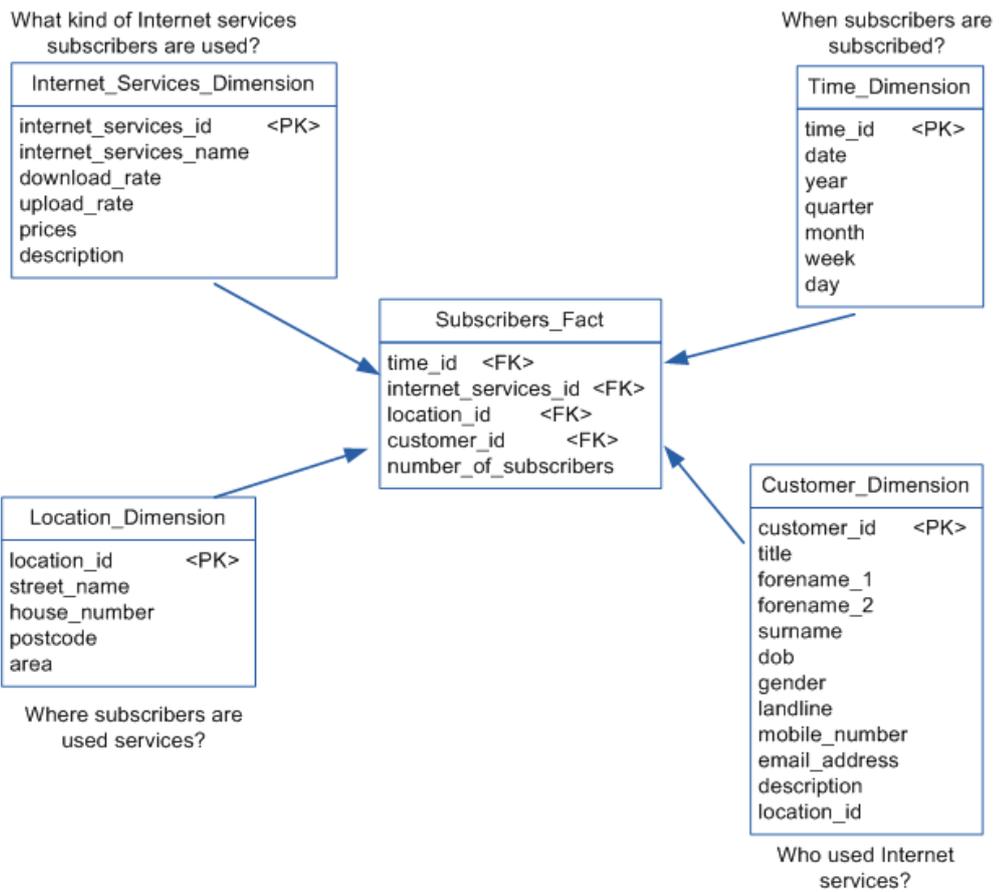

Figure 3-7: Demonstration of simple star diagram for the Gorannet Company

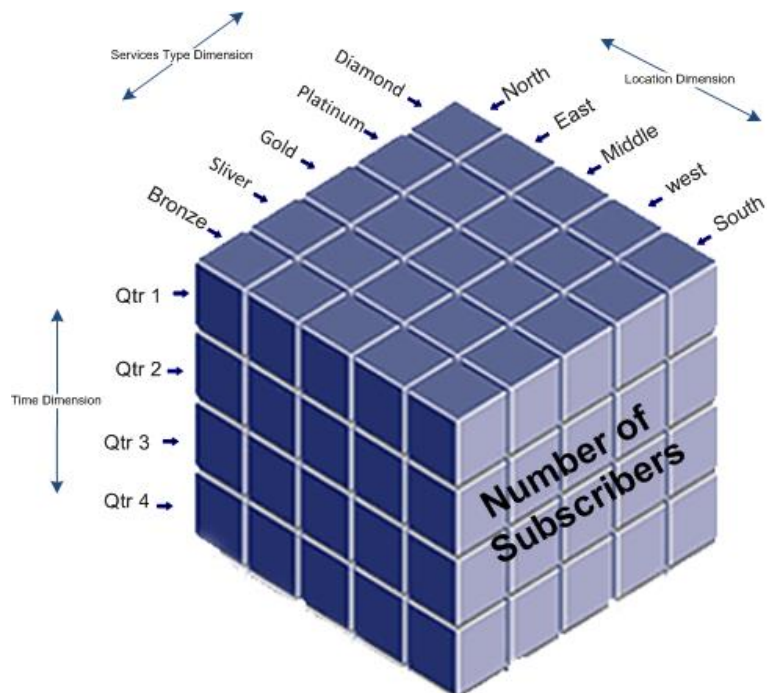

Figure 3-8: Gorannet Data Cube, multidimensional model logical view

# 4. RESULTS AND DISCUSSION

There is limited research information on the topic of multidimensional database. One piece of relative research on the subject matter is provided by (Manisarma Vittapu et al., 2011). Relative to the information gathered in this paper, an argument exists for a framework that is introduced for which design methodologies are proposed to map the relational database into a multidimensional model. This process is described to permeate the relational database into metrics and dimensional attributes by applying the proposed set of mapping rules. However, the framework is designed in the most simplified manner so that it can be applied to in complex databases and cannot adhere to facilitate the multidimensional database authorities since it does not maintain numerous objectives and benefits of multidimensional databases. With that said, the objective of this research is to introduce a new framework of information on the subject of multidimensional database and propose a solution that will make using multidimensional databases, easier and convenient for both experts and non-experts alike.

## 4.1 The New Framework

As it has been mentioned before, the proposed framework is based on the performance of a relational database. For that reason, obtaining the data source potentially differs depending on the contributing factors when constructing a query. The popular ways to import the relational database in order to reverse engineering is as follows:

- From data dictionary.
- From data definition language (DDL) files.

The proposed framework initially shares a similar beginning to other projects both in literature and experiences within industrial projects. First and foremost, a prototype provides a certain functionality and set of data. This prototype is further adopted according to the altering and developing requirements obtained from users' feedback.

A star schema was preferred for the proposed framework because it is the most utilised schema template when implementing the OLAP system. More importantly, the utilisation of the star schema alleviates the query process for users through a set of convenient steps on how to create a fact dimension table based on prior concerns when producing data analysis. It is important to note that user requirements may vary when using data warehouse maintenance depending on the type of query.

In regards to the design of conceptual multidimensional schema, it should be specified on a conceptual level to satisfy the flexibility and reusability of the schema. This means that, it should not assume any facts that are the result of further design steps. Moreover, the logical multidimensional model is implemented through a physical schema.

The design process can be dynamic because of the approaches that follow when using multidimensional databases projects. One such approach is through obtaining certain requirements from the relational schema for which the interactive multidimensional analysis technology is based upon. Moreover, businesses decisions are influenced by the production of data analysis. Therefore, it is important for the multidimensional technology to adapt and accommodate to all analysis requirements. Data analysis is a critical success factor for effectively using the multidimensional technology because it surrounds the elements that are necessary to meet the needs of analysts. During the requirement analysis, an array of queries may require different data while the multidimensional schema determines the possible analysis capabilities and these changes lead to an altered multidimensional database schema. Figure 4-1 illustrates the overall schema design process based on the user requirements and OLTP schema.

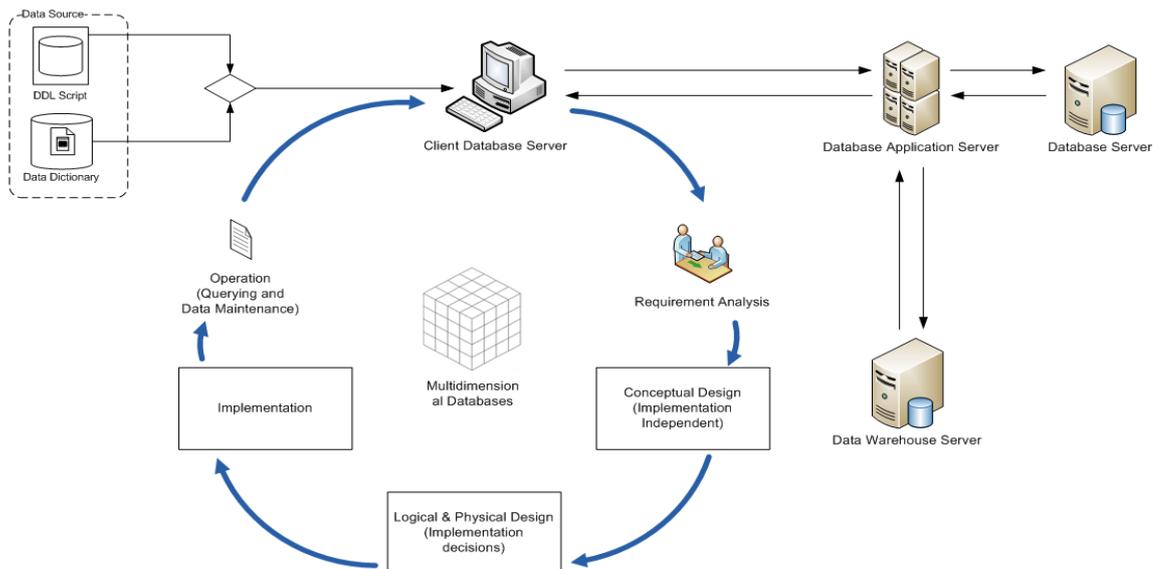
Figure 4-1: Diagram of multidimensional database schema design process and maintenance cycle

- **Identifying user requirements:** The requirement analysis of users might consist of a data scope; granularity, structure and quality are gathered to improved assess the informational needs of users. Consequently, the result of analytical requirements is a set of multidimensional views that have to be supported by the information system.
- **Conceptual design:** Based on the particular user requirements, the conceptual design is produced. This objective is used to consolidate the required views into a single conceptual model. The multidimensional database schema that was developed is based on identifying user requirements. This sets up the main stage of the data modelling process and it prepares for the next iterations.
- **Logical and physical design:** This stage consists of the implementation decision phase, where iteration is optimised and an identity measure occurs. If necessary, the relational tables should be demoralised.
- **Implementation:** After the logical and physical design has developed, the implementation process for developing schemas during the technical design phase occurs. Furthermore, ETL operations can be performed in this step; loading the data into one of the multidimensional database for instance, star schema.
- **Operation:** Following the implementation phase, the data can be found in the OLAP database. Thereafter, it should be tested in order to meet the user requirements. During the first iteration, if the user requirements are met, a new cycle is started. Most likely, it needs to repeat one or more iterations of the cycle process to arrive at the ideal multidimensional database schema.

## 4.2 Case Studies Evaluations
During the process of applying the two case studies in the proposed framework that is converting relational database to data warehouse schema or star schema. Measure attributes in the fact table depend on all foreign keys in dimension tables. The measure attributes value represents a business values (business facts). Business facts are the converted vales from any selected fields or descriptive attributes in the relational tables to meaningful vales.  For instance, in the Gorannet company scenario, the company subscriber's fact table is represented by cubes whereby the cubes contain measures which inherit its form. Furthermore, the measures share characteristics among other measures in the cube.

While creating a dimension table, the dimension descriptive attributes should have a clear label name and complete data to provide quality assurance. It is vital that no data is missing from the descriptive attributes for best results.

The idea of using simple integer vales (foreign keys and primary keys) serves to connect fields between fact tables and dimension tables. This connection is referred to as surrogate keys. Additionally, the simple integer vales become meaningless despite advantageous characteristics such as performance, updating, and easily changing tracks in dimension vales. The advantage of having surrogate keys serves to maintain information in a data warehouse as a dimension table is changed.

The essential aim of dimension tables is to construct standardised, adjusted dimensions that can be represented differently for business processing. The adjusted dimensions are crucial elements because it promotes consistency, integration, and reduced development time to market.

## 5. CONCLUSION

This research serves as a contribution to the on-going research in the field of multidimensional databases in a way that proposes a novel approach to design a new framework to adopt multidimensional databases for organizational information system strategies. Mainly, this study has focused on four points. Firstly, proposes a new framework for adopting multidimensional databases from a logical schema that is built in a relational database. Secondly, it introduces a proposition for a new framework that is more organized and efficient to utilise. The framework would alleviate the decision making process and improve the organisational informational system strategies. Thirdly, the new framework allows for the creation of the semi-automatic method by supporting multidimensional database. The two essential reasons for designing the framework were to provide user with simplistic instructions and tools to build individual multidimensional databases. Furthermore, the proposed framework does not intend to deter businesses from using existing methods. Rather, it seeks to alleviate the technical concerns of non-experts and provide managers with a simple model that can be used by small or medium organisations. Finally, the proposed framework demonstrates various concepts familiar to OLAP users, such as dimensions, levels, paths, measures and cubes. Each of them is interpreted alongside a detailed illustration. These components are explained in order to enhance the advantageous and process of the proposed framework. In brief, the overall objective for proposing a new conceptual multidimensional database is to provide an expeditious method for retrieving definitive information from relational databases that includes a multidimensional database technique already utilised for aiding the decision making process.

The main future works of the proposed framework are presented as follows: Firstly, this framework can be applied to the Gorannet Company by first creating a data warehouse project and then converting the OLTP system to an OLAP system by corresponding logical evolution capabilities for star schemas in order to access the data in a better way and identity the overall company's business trends. Secondly, the proposed framework method is based on relational database in organisations. The framework could be extended to a web based application project.

**Acknowledgment**

This study has been done in Kurdistan institution for strategic studies and scientific research and Sulaimani Governorate. The authors wish to express their deep thanks to Professor Dr Polla Khanaqa, the head of the Kurdistan Institution; Dr Aso Faraydun, the governor of Sulaimani Province; Mr Abdulkhalikh Muhammed, the general administrative manager in Sulaimani Governorate; and Mr Hunar Tuafiq, an advisor of Sulaimani governor for their kind support in conducting this study.

**References**

AGRUSA, R. L., BURIAN, J., JEDLICKA, J., GRIESSL, R., LEDERBUCH, P. & PROCHAZKA, M. 2012. Operational process control data server. Google Patents.

CHRISTIAN, S. J. & TORBEN BACH, P. 2007. Multidimensional Databases and OLAP. 53, 53-1,53-27.

DASH, A. K. & AGARWAL, R. 2001. Dimensional modeling for a data warehouse. ACM SIGSOFT software engineering notes, 26, 83-84.

EL-SAPPAGH, S. H. A., HENDAWI, A. M. A. & EL BASTAWISSY, A. H. 2011. A proposed model for data warehouse ETL processes. Journal of King Saud University-Computer and Information Sciences, 23, 91-104.

FERNANDEZ ORTEGA, A. 2013. Compensating for unbalanced hierarchies when generating OLAP queries from report specifications. US Patent 8,484,157.

GAO, B., ZHANG, S. & YAO, N. A Multidimensional Pivot Table Model Based on MVVM Pattern for Rich Internet Application. Computer, Consumer and Control (IS3C), 2012 International Symposium on, 2012. IEEE, 24-27.

GUPTA, P. 2013. Data Warehouse and Data Mining Technology -A study of its impact, relevance and need in Enterprises of Delhi NCR region. IJRIT International Journal of Research in Information Technology, 1, 37-50.

INMON, W. 2000. WHAT IS AData WAREHOUSE?

JENSEN, C. S., PEDERSEN, T. B. & THOMSEN, C. 2010. Multidimensional databases and data warehousing. Synthesis Lectures on Data Management, 2, 1-111.

KANTARDZIC, M. 2011. Data mining: concepts, models, methods, and algorithms, John Wiley & Sons.

KARAYANNIDIS, N., TSOIS, A., SELLIS, T., PIERINGER, R., MARKL, V., RAMSAK, F., FENK, R., ELHARDT, K. & BAYER, R. Processing star queries on hierarchically-clustered fact tables. Proceedings of the 28th international conference on Very Large Data Bases, 2002. VLDB Endowment, 730-741.

KIMBALL, R. 1998. The data warehouse lifecycle toolkit: expert methods for designing, developing, and deploying data warehouses, Wiley. com.

LECHTENBÖRGER, J. & VOSSEN, G. 2003. Multidimensional normal forms for data warehouse design. Information Systems, 28, 415-434.

MANISARMA VITTAPU, RAMBABU DESODU & PREMCHAND P, P. R. V. 2011. A New User Framework to Design MD Model from Relational One. Computer Applications (JCA), Volume IV (4), 127-130.

NELSON, G. S. 2008. Building OLAP Cubes with SAS 9: A Hands on Workshop. Wave Technologies, 2.

PEDERSEN, T. B. & JENSEN, C. S. 2005. Multidimensional Databases. Citeseer.

PUJOLLE, G., RAVAT, F., TESTE, O., TOURNIER, R. & ZURFLUH, G. 2011. Multidimensional database design from document-centric XML documents. Data Warehousing and Knowledge Discovery. Springer.

ZAMAN, K. A., SONG, S. & SUEN, E. S.-L. 2010. Conversion of a relational database query to a query of a multidimensional data source by modeling the multidimensional data source. Google Patents.